\documentclass[12pt]{article}
\usepackage{amsmath}
\usepackage{amssymb}

\title{
Exceptional orthogonal polynomials, exactly solvable potentials and supersymmetry}
\author{C Quesne\\ 
{\small Physique Nucl\'eaire Th\'eorique et Physique Math\'ematique,  Universit\'e Libre de Bruxelles,} \\ 
{\small Campus de la Plaine CP229, Boulevard~du Triomphe, B-1050 Brussels, Belgium}}
\date{ }
\begin{document}
\baselineskip=22pt plus 1pt minus 1pt
\maketitle

\begin{abstract} 
We construct two new exactly solvable potentials giving rise to bound-state solutions to the Schr\"odinger equation, which can be written in terms of the recently introduced Laguerre- or Jacobi-type $X_1$ exceptional orthogonal polynomials. These potentials, extending either the radial oscillator or the Scarf I potential by the addition of some rational terms, turn out to be translationally shape invariant as their standard counterparts and isospectral to them.
\end{abstract}

\noindent
Short title: Exactly solvable potentials 


%
%
\newpage
Classical orthogonal polynomials are known to play a fundamental role in the construction of bound-state solutions to exactly solvable potentials in quantum mechanics. For such a purpose, the factorization method~\cite{schrodinger, infeld} and its realization in supersymmetric quantum mechanics (SUSYQM)~\cite{witten}, especially for shape-invariant potentials~\cite{gendenshtein}, as well as the equivalent Darboux transformation~\cite{darboux}, prove very useful. The same is true for a more traditional approach, the point canonical transformation (PCT) method~\cite{bhatta}, consisting in directly mapping Schr\"odinger equations into the second-order differential equations satisfied by those polynomials.\par
%
%
On the other hand, bound-state solutions to exactly solvable potentials are by no way restricted to classical orthogonal polynomials. For instance, SUSYQM~\cite{sukumar, cooper, junker, bagchi00} and the Darboux transformation~\cite{bagrov, bagchi98, gomez04} are very efficient at producing new sophisticated exactly solvable potentials by adding or deleting some states, or else by leaving the spectrum unchanged. The PCT method is also very powerful for generating new shape-invariant or non-shape-invariant potentials not only in a standard context~\cite{levai}, but also in more general ones, such as those of quasi-exact~\cite{bagchi03} or conditionally-exact~\cite{roychoudhury} solvability and that of position-dependent masses~\cite{bagchi05}.\par
%
%
Very recently, two new families of exceptional orthogonal $n$th-degree polynomials, $\hat{P}^{(\alpha,\beta)}_n(x)$ and $\hat{L}^{(\alpha)}_n(x)$, $n=1$, 2, 3,~\ldots, have been introduced~\cite{gomez08a, gomez08b}. Such sequences, referred to as Jacobi- or Laguerre-type $X_1$ polynomials, respectively, arise as solutions of second-order eigenvalue equations with rational coefficients. They are characterized by the remarkable property that although they do not start with a constant but with a linear polynomial, they form complete sets with respect to some positive-definite measure in contrast with what would happen if one deleted their first member from the families of classical orthogonal polynomials.\par
%
%
In this communication, we plan to show that there exist some exactly solvable potentials whose bound-state wavefunctions can be written in terms of these new exceptional orthogonal polynomials. For such a purpose, we shall make use of the standard PCT method. In a second step, we shall employ SUSYQM techniques to prove that our new exactly solvable potentials are transitionally shape invariant.\par
%
%
In the PCT method, one looks for solutions of the Schr\"odinger equation,
\begin{equation}
  H \psi(x) \equiv \left(- \frac{d^2}{dx^2} + V(x)\right) \psi(x) = E \psi(x),  \label{eq:SE}
\end{equation}
of the form
\begin{equation}
  \psi(x) = f(x) F(g(x)),  \label{eq:psi}
\end{equation}
where $f(x)$, $g(x)$ are two so far undetermined functions and $F(g)$ satisfies a second-order differential equation
\begin{equation}
  \ddot{F} + Q(g) \dot{F} + R(g) F = 0.  \label{eq:eq-F}
\end{equation}
Here a dot denotes derivative with respect to $g$.\par
%
%
On inserting equation (\ref{eq:psi}) in equation (\ref{eq:SE}) and comparing the result with equation (\ref{eq:eq-F}), one arrives at two expressions for $Q(g(x))$ and $R(g(x))$ in terms of $E - V(x)$ and of $f(x)$, $g(x)$ and their derivatives. The former allows one to calculate $f(x)$, which is given by
\begin{equation}
  f(x) \propto \frac{1}{\sqrt{g'}} \exp\left(\frac{1}{2} \int^{g(x)} Q(u)\, du\right),  \label{eq:f}
\end{equation}
while the latter leads to the equation
\begin{equation}
  E - V(x) = \frac{g'''}{2g'} - \frac{3}{4} \left(\frac{g''}{g'}\right)^2 + g^{\prime2} \left(R - \frac{1}{2} 
  \dot{Q} - \frac{1}{4} Q^2\right).  \label{eq:PCT}
\end{equation}
In (\ref{eq:f}) and (\ref{eq:PCT}), a prime denotes derivative with respect to $x$. For equation (\ref{eq:PCT}) to be satisfied, one needs to find some function $g(x)$ ensuring the presence of a constant term on its right-hand side to compensate $E$ on its left-hand one, while giving rise to a potential $V(x)$ with well-behaved
wavefunctions.\par
%
%
Let us start by considering for the second-order differential equation (\ref{eq:eq-F}) that satisfied by Laguerre-type $X_1$ polynomials $\hat{L}^{(\alpha)}_n(x)$, $n=1$, 2, 3,~\ldots, $\alpha > 0$. In such a case, the functions $Q(g)$ and $R(g)$ can be expressed as~\cite{gomez08a, gomez08b}
\begin{equation*}
\begin{split}
  Q(g) &= - \frac{(g-\alpha)(g+\alpha+1)}{g(g+\alpha)} = - 1 + \frac{\alpha +1}{g} - \frac{2}{g+\alpha}, \\
  R(g) &= \frac{1}{g} \left(\frac{g-\alpha}{g+\alpha} + n - 1\right) = \frac{n-2}{g} + \frac{2}{g+\alpha},
\end{split}
\end{equation*}
so that we obtain
\begin{equation}
  R - \frac{1}{2} \dot{Q} - \frac{1}{4} Q^2 = - \frac{1}{4} + \frac{2\alpha n + \alpha^2 - \alpha + 2}
  {2\alpha g} - \frac{1}{\alpha(g+\alpha)} - \frac{(\alpha+1)(\alpha-1)}{4g^2} - \frac{2}{(g+\alpha)^2}.
  \label{eq:Laguerre}
\end{equation}
\par
%
%
A constant term can be generated on the right-hand side of equation (\ref{eq:PCT}) by assuming $g^{\prime2}/g = C$, which can be achieved by taking $g(x) = \frac{1}{4} C x^2$. Equations (\ref{eq:PCT}) and (\ref{eq:Laguerre}) then yield
\begin{equation*}
\begin{split}
  E &= \frac{1}{2} C (2n + \alpha - 1), \\
  V(x) &= \frac{1}{16} C^2 x^2 + \frac{(\alpha - \frac{1}{2})(\alpha + \frac{1}{2})}{x^2} + \frac{4C}{C x^2
         + 4\alpha} - \frac{32 C \alpha}{(C x^2 + 4\alpha)^2}.
\end{split}
\end{equation*}
On setting
\begin{equation*}
  C = 2\omega, \qquad \alpha = l + \tfrac{1}{2}, \qquad n = \nu + 1,
\end{equation*}
we arrive at
\begin{equation}
  E_{\nu} = \omega (2\nu + l + \tfrac{3}{2}), \qquad \nu=0, 1, 2, \ldots,  \label{eq:E-L}
\end{equation}
and
\begin{equation}
\begin{split}
  V(x) &= V_1(x) + V_2(x), \\ 
  V_1(x) &= \frac{1}{4} \omega^2 x^2 + \frac{l(l+1)}{x^2}, \\
  V_2(x) &= \frac{4\omega}{\omega x^2 + 2l + 1} - \frac{8\omega (2l+1)}{(\omega x^2 + 2l + 1)^2}. 
\end{split}  \label{eq:V-L}
\end{equation}
\par
%
%
{}For 
\begin{equation*}
  0 < x < \infty, \qquad \omega > 0, \qquad l = 0, 1, 2, \ldots,
\end{equation*}
$V(x)$ is a well-behaved potential, which may be interpreted as an $l$-dependent (effective) potential, extending the standard radial oscillator potential $V_1(x)$ by the addition of some rational terms. Such terms do not change the behaviour of the conventional potential for large values of $x$, while for small values they have only a drastic effect when the angular momentum $l$ vanishes, in which case $V(0) = - 4 \omega < 0$ instead of $V(0) = 0$.\par
%
%
{}From equation (\ref{eq:E-L}), it follows that the extended potential has the same spectrum as the standard one. The corresponding wavefunctions can be found from equations (\ref{eq:psi}) and (\ref{eq:f}). On solving the latter for the choices made here for $Q(g)$ and $g(x)$, we get
\begin{equation*}
  \psi_{\nu}(x) = {\cal N}_{\nu} \frac{x^{l+1}}{\omega x^2 + 2l + 1} \hat{L}^{\left(l + \frac{1}{2}\right)}
  _{\nu+1}(\tfrac{1}{2} \omega x^2) e^{- \frac{1}{4} \omega x^2},
\end{equation*}
where the normalization constant is obtained from equations (31), (33) and (34) of \cite{gomez08a} as
\begin{equation*}
  {\cal N}_{\nu} = \left(\frac{\omega^{l + \frac{3}{2}} \nu!}{2^{l - \frac{3}{2}} \left(\nu + l + \frac{3}{2}
  \right) \Gamma\left(\nu + l + \frac{1}{2}\right)}\right)^{1/2}.
\end{equation*}
\par
%
%
In particular, the ground-state wavefunction can be written as
\begin{equation}
  \psi_0(x) \propto \psi_{10}(x) [1 + \phi(x)], \qquad \psi_{10}(x) \propto x^{l+1} e^{- \frac{1}{4} \omega
   x^2}, \qquad \phi(x) = \frac{2}{\omega x^2 + 2l + 1},  \label{eq:gs-L}
\end{equation}
and differs from that of the standard radial oscillator, $\psi_{10}(x)$, by the extra factor $1 + \phi(x)$. It is obvious that it is a zero-node function on the half-line, as it shoud be. Furthermore, it can be easily checked by direct calculation that it satisfies equation (\ref{eq:SE}) for the potential (\ref{eq:V-L}) and $E_0 = \omega \left(l + \frac{3}{2}\right)$.\par
%
%
More generally, as shown in \cite{gomez08b}, the polynomial $\hat{L}^{\left(l + \frac{1}{2}\right)}_{\nu+1} (\tfrac{1}{2} \omega x^2)$ (and hence the wavefunction $\psi_{\nu}(x)$) has $\nu$ zeroes on the half-line. From general properties of the one-dimensional Schr\"odinger equation, it therefore results that we have found all the eigenvalues of potential (\ref{eq:V-L}).\par
%
%
Let us next consider the case where the second-order differential equation (\ref{eq:eq-F}) coincides with that satisfied by Jacobi-type $X_1$ polynomials $\hat{P}^{(\alpha,\beta)}_n(x)$, $n=1$, 2, 3,~\ldots, $\alpha, \beta > -1$, $\alpha \ne \beta$ \cite{gomez08a, gomez08b}, i.e.,
\begin{equation*}
\begin{split}
  Q(g) &= - \frac{(\beta+\alpha+2) g - (\beta-\alpha)}{1-g^2} - \frac{2(\beta-\alpha)}{(\beta-\alpha) g -
        (\beta+\alpha)}, \\
  R(g) &= - \frac{(\beta-\alpha) g - (n-1) (n+\beta+\alpha)}{1-g^2} - \frac{(\beta-\alpha)^2}{(\beta-\alpha) g
        -(\beta+\alpha)}.
\end{split}
\end{equation*}
This choice leads to
\begin{equation*}
  R - \frac{1}{2} \dot{Q} - \frac{1}{4} Q^2 = \frac{Cg+D}{1-g^2} + \frac{Gg+J}{(1-g^2)^2} + \frac{K}{(\beta
  -\alpha) g - (\beta+\alpha)} + \frac{L}{[(\beta-\alpha) g - (\beta+\alpha)]^2}, 
\end{equation*}
where
\begin{equation*}
\begin{split}
  C &= \frac{(\beta-\alpha)(\beta+\alpha)}{2\alpha\beta}, \quad D = n^2 + (\beta+\alpha-1) n + \frac{1}{4}
        [(\beta+\alpha)^2 - 2(\beta+\alpha) - 4] + \frac{\beta^2 + \alpha^2}{2\alpha\beta}, \\
  G &= \frac{1}{2}(\beta-\alpha)(\beta+\alpha), \quad J = - \frac{1}{2}(\beta^2 + \alpha^2 - 2), \\
  K &= \frac{(\beta-\alpha)^2 (\beta+\alpha)}{2\alpha\beta}, \quad L = - 2(\beta-\alpha)^2.
\end{split}
\end{equation*}
\par
%
%
We can obtain a constant term on the right-hand side of (\ref{eq:PCT}) by assuming $g^{\prime2}/(1-g^2) = \bar{C}$. For $\bar{C} = a^2 > 0$, we can take $g(x) = \sin(ax)$. On rescaling the variable $x$, the parameter $a$ can be set equal to one. Then with the changes of parameters and of quantum number
\begin{align*}
  \alpha &= A - B - \tfrac{1}{2}, \qquad \beta = A + B - \tfrac{1}{2} \qquad \mbox{or} \qquad A = \tfrac{1}{2}
       (\beta+\alpha+1), \qquad B = \tfrac{1}{2}(\beta-\alpha), \\
  n &= \nu + 1,
\end{align*}
we arrive at the following results
\begin{equation}
  E_{\nu} = (\nu + A)^2, \qquad \nu = 0, 1, 2, \ldots,  \label{eq:E-J}
\end{equation}
and
\begin{equation}
\begin{split}
  V(x) &= V_1(x) + V_2(x), \\
  V_1(x) &= [A(A-1) + B^2] \sec^2 x - B(2A-1) \sec x \tan x, \\
  V_2(x) &= \frac{2(2A-1)}{2A-1 - 2B \sin x} - \frac{2[(2A-1)^2 - 4B^2]}{(2A-1 - 2B \sin x)^2}. 
\end{split}  \label{eq:V-J}
\end{equation}
\par
%
%
The function $V_1(x)$ defines a Scarf I potential, for which it is customary to assume
\begin{equation*}
  - \frac{\pi}{2} < x < \frac{\pi}{2}, \qquad 0 < B < A-1.
\end{equation*}
For such values of the variable and of the parameters, the full potential $V(x)$ has the same behaviour as $V_1(x)$ for $x \to \pm \pi/2$, only the position and the value of the minimum being modified.\par
%
%
It results from (\ref{eq:E-J}) that on using as usual Dirichlet boundary conditions at the end points of the interval, the extended Scarf I potential (\ref{eq:V-J}) has the same spectrum as the conventional one. Its wavefunctions can be written as
\begin{equation*}
  \psi_{\nu}(x) = {\cal N}_{\nu} \frac{(1 - \sin x)^{\frac{1}{2}(A-B)} (1 + \sin x)^{\frac{1}{2}(A+B)}}
  {2A-1 - 2B \sin x} \hat{P}^{\left(A-B-\frac{1}{2}, A+B-\frac{1}{2}\right)}_{\nu+1}(\sin x),
\end{equation*}
where the normalization constant
\begin{equation*}
  {\cal N}_{\nu} = \frac{B}{2^{A-2}} \left(\frac{\nu!\, (2\nu+2A) \Gamma(\nu+2A)}
  {\left(\nu+A-B+\frac{1}{2}\right) \left(\nu+A+B+\frac{1}{2}\right) \Gamma\left(\nu+A-B-\frac{1}{2}\right)
  \Gamma\left(\nu+A+B-\frac{1}{2}\right)}\right)^{1/2}
\end{equation*}
is a consequence of equations (23), (25) and (26) of \cite{gomez08a}.\par
%
%
The ground-state wavefunction assumes the simple form
\begin{equation}
\begin{split}
  \psi_0(x) &\propto \psi_{10}(x)[1 + \phi(x)], \qquad \psi_{10}(x) \propto (1 - \sin x)^{\frac{1}{2}(A-B)} 
         (1 + \sin x)^{\frac{1}{2}(A+B)}, \\
  \phi(x) &= \frac{2}{2A-1 - 2B \sin x},
\end{split}  \label{eq:gs-J}
\end{equation}
where the presence of $\phi(x)$ is due to the rational terms in $V_2(x)$. As in the previous case, the function $\psi_0(x)$ has no node on the interval of variation of $x$ and can be easily checked to satisfy equation (\ref{eq:SE}).\par
%
%
More generally, the polynomial $\hat{P}^{\left(A-B-\frac{1}{2}, A+B-\frac{1}{2}\right)}_{\nu+1}(\sin x)$ (and hence the wavefunction $\psi_{\nu}(x)$) has $\nu$ zeroes on $\left(- \frac{\pi}{2}, \frac{\pi}{2}\right)$ \cite{gomez08b}, so that no other eigenvalues than (\ref{eq:E-J}) may exist for potential (\ref{eq:V-J}).\par
%
%
Let us finally combine our results with SUSYQM methods \cite{sukumar, cooper, junker, bagchi00}. In the minimal version of SUSY, the supercharges $Q$ and $Q^{\dagger}$ are generally assumed to be represented by $Q = A \sigma_-$, $Q^{\dagger} = A^{\dagger} \sigma_+$, where $\sigma_{\pm}$ are combinations $\sigma_{\pm} = \sigma_1 \pm {\rm i} \sigma_2$ of the Pauli matrices and $A$, $A^{\dagger}$ are taken to be first-derivative differential operators, $A = \frac{d}{dx} + W(x)$, $A^{\dagger} = - \frac{d}{dx} + W(x)$, with $W(x)$ known as the superpotential. The supersymmetric Hamiltonian $H_{\rm s} = \{Q, Q^{\dagger}\}$ is diagonal, i.e., $H_{\rm s} = \operatorname{diag} (H^{(+)}, H^{(-)})$, and its components $H^{(\pm)}$ can be written in factorized form in terms of $A$ and $A^{\dagger}$,
\begin{equation*}
  H^{(+)} = A^{\dagger} A = - \frac{d^2}{dx^2} + V^{(+)}(x) - E, \qquad H^{(-)} = A A^{\dagger} = 
  - \frac{d^2}{dx^2} + V^{(-)}(x) - E, 
\end{equation*}
at some arbitrary factorization energy $E$. The partner potentials $V^{(\pm)}(x)$ are related to $W(x)$ through $V^{(\pm)}(x) = W^2(x) \mp W'(x) + E$.\par
%
%
The spectrum of $H_{\rm s}$ is doubly degenerate except possibly for the ground state. In the exact SUSY case to be considered here, the ground state at vanishing energy is nondegenerate. In the present notational set-up, it belongs to $H^{(+)}$,
\begin{equation*}
  H^{(+)} \psi^{(+)}_0(x) = 0, \qquad \psi^{(+)}_0(x) \propto \exp \left(- \int^x W(t) dt\right).
\end{equation*}
\par
%
%
Let us identify $V^{(+)}(x)$ with either the extended radial oscillator potential (\ref{eq:V-L}) or the extended Scarf I potential (\ref{eq:V-J}) and take for the factorization energy $E_0 = \omega (l + \frac{3}{2})$ or $E_0 = A^2$, respectively. Then $\psi^{(+)}_0(x)$ is given by equation (\ref{eq:gs-L}) or equation (\ref{eq:gs-J}). The corresponding superpotential $W(x) = - \psi'_0(x)/\psi_0(x)$ can be separated into two parts,
\begin{equation*}
  W(x) = W_1(x) + W_2(x), \qquad W_1(x) = - \frac{\psi'_{10}}{\psi_{10}}, \qquad W_2(x) = - 
  \frac{\phi'}{1+\phi},
\end{equation*}
where $W_1(x)$ is the superpotential for the conventional potential, i.e.,
\begin{equation*}
  W_1(x) = \frac{1}{2} \omega x - \frac{l+1}{x} \qquad \mbox{\rm or} \qquad W_1(x) = A \tan x - B \sec x,
\end{equation*}
and the additional term $W_2(x)$ can be written as
\begin{equation*}
  W_2(x) = 2\omega x \left(\frac{1}{\omega x^2 + 2l + 1} - \frac{1}{\omega x^2 + 2l + 3}\right)
\end{equation*}
or 
\begin{equation*}
  W_2(x) = - 2B \cos x \left(\frac{1}{2A-1 - 2B \sin x} - \frac{1}{2A+1 - 2B \sin x} \right), 
\end{equation*}
respectively.\par
%
%
{}From this, it follows that the partner potential $V^{(-)}(x)$ to $V^{(+)}(x)$ (with one less eigenvalue) is given by
\begin{equation*}
  V^{(-)}(x) = V^{(+)}(x) + 2W'(x) = V^{(-)}_1(x) + V^{(-)}_2(x), \quad V^{(-)}_i(x) = V^{(+)}_i(x) + 
  2W'_i(x), \quad i = 1, 2.
\end{equation*}
As it is well known, $V_1^{(-)}(x)$ is a standard radial oscillator (resp.\ Scarf I) potential with $l$ replaced by $l+1$ (resp.\ $A$ replaced by $A+1$). It is straightforward to convince oneself that a similar property relates $V_2^{(-)}(x)$ with $V_2^{(+)}(x)$. We therefore conclude that the two extended potentials (\ref{eq:V-L}) and (\ref{eq:V-J}) are translationally shape invariant as their conventional counterparts.\par
%
%
Summarizing, we have constructed some exactly solvable potentials for which the recently introduced Laguerre- or Jacobi-type $X_1$ exceptional orthogonal polynomials play a fundamental role. Furthermore, we have demonstrated that these new potentials are shape invariant. It is rather obvious that the method described here could be used for other choices of function $g(x)$ in order to generate other types of potentials connected with such polynomials. Another interesting open question for future work would be the origin of the (strict) isospectrality observed between the extended and conventional potentials.\par
%
%


\newpage
\begin{thebibliography}{99}

\bibitem{schrodinger} Schr\"odinger E 1940 {\sl Proc.\ R.\ Irish Acad.} A {\bf 46} 9, 183 \\
Schr\"odinger E 1941 {\sl Proc.\ R.\ Irish Acad.} A {\bf 47} 53

\bibitem{infeld} Infeld L and Hull T E 1951 {\sl Rev.\ Mod.\ Phys.} {\bf 23} 21

\bibitem{witten} Witten E 1981 {\sl Nucl.\ Phys.} B {\bf 188} 513

\bibitem{gendenshtein} Gendenshtein L E 1983 {\sl JETP Lett.} {\bf 38} 356

\bibitem{darboux} Darboux G 1888 {\sl Th\'eorie G\'en\'erale des Surfaces} vol 2 (Paris: Gauthier-Villars)

\bibitem{bhatta} Bhattacharjie A and Sudarshan E C G 1962 {\sl Nuovo Cimento} {\bf 25} 864

\bibitem{sukumar} Sukumar C V 1985 {J.\ Phys.\ A: Math.\ Gen.} {\bf 18} 2917

\bibitem{cooper} Cooper F, Khare A and Sukhatme U 1995 {\sl Phys.\ Rep.} {\bf 251} 267

\bibitem{junker} Junker G 1996 {\sl Supersymmetric Methods in Quantum and Statistical Physics} (Berlin: Springer)

\bibitem{bagchi00} Bagchi B 2000 {\sl Supersymmetry in Quantum and Classical Physics} (Boca Raton, FL: Chapman and Hall/CRC)

\bibitem{bagrov} Bagrov V G and Samsonov B F 1995 {\sl Theor.\ Math.\ Phys.} {\bf 104} 1051

\bibitem{bagchi98} Bagchi B and Ganguly A 1998 {\sl Int.\ J.\ Mod.\ Phys.} A {\bf 13} 3711

\bibitem{gomez04} G\'omez-Ullate D, Kamran N and Milson R 2004 {\sl J.\ Phys.\ A: Math.\ Gen.} {\bf 37} 1789, 10065

\bibitem{natanzon} Natanzon G 1979 {\sl Theor.\ Math.\ Phys.} {\bf 38} 146

\bibitem{levai} L\'evai G 1989 {\sl J.\ Phys.\ A: Math.\ Gen.} {\bf 22} 689 \\
L\'evai G 1991 {\sl J.\ Phys.\ A: Math.\ Gen.} {\bf 24} 131

\bibitem{bagchi03} Bagchi B and Ganguly A 2003 {\sl J.\ Phys.\ A: Math.\ Gen.} {\bf 36} L161

\bibitem{roychoudhury} Roychoudhury R, Roy P, Znojil M and L\'evai G 2001 {\sl J.\ Math.\ Phys.} {\bf 42} 1996

\bibitem{bagchi05} Bagchi B, Gorain P, Quesne C and Roychoudhury R 2005 {\sl Europhys.\ Lett.} {\bf 72} 155

\bibitem{gomez08a} G\'omez-Ullate D, Kamran N and Milson R 2008 An extension of Bochner's problem: exceptional invariant subspaces {\sl Preprint} math-ph/0805.3376

\bibitem{gomez08b} G\'omez-Ullate D, Kamran N and Milson R 2008 An extended class of orthogonal polynomials defined by a Sturm-Liouville problem {\sl Preprint} math-ph/0807.3939

\end {thebibliography}
\end{document}